%% file: main.tex
\documentclass[fleqn,10pt]{wlscirep}
\usepackage[utf8]{inputenc}
\usepackage[T1]{fontenc}

\usepackage{subfiles}
\usepackage{multirow}
\usepackage{tikz} % 提供对号 \checkmark

\usepackage{amsthm,amsmath,amssymb}
\usepackage{mathptmx}
\usepackage{mathrsfs}
\DeclareMathAlphabet{\mathcal}{OMS}{cmsy}{m}{n}  % 区分\mathcal,\mathscr

\definecolor{cred}{HTML}{FF6B6B}
\definecolor{cyellow}{HTML}{FEC260}
\definecolor{cgreen}{HTML}{6BCB77}
\definecolor{cgreen}{HTML}{70AD47}
\definecolor{cblue}{HTML}{4D96FF}
\definecolor{cpurple}{HTML}{2A0944}
\definecolor{ggray}{RGB}{127,127,127}
\definecolor{aliceblue}{rgb}{0.94, 0.97, 1.0}

\usepackage{makecell}   % 表格内换行

\title{
 Deep peak property learning for efficient chiral molecules ECD spectra prediction
}

\author[1,2,4*]{Hao Li}
\author[3*]{Da Long}
\author[1,2,4$\dagger$]{Li Yuan}
\author[1,2,4]{Yonghong Tian}
\author[3$\dagger$]{Xinchang Wang}
\author[2,5$\dagger$]{Fanyang Mo}
\affil[1]{\textit{School of Electronic and Computer Engineering, Peking University, Shenzhen, China}}
\affil[2]{\textit{AI for Science (AI4S)-Preferred Program, Peking University Shenzhen Graduate School, Shenzhen, China}}
\affil[3]{\textit{State Key Laboratory of Physical Chemistry of Solid Surfaces, School of Electronic Science and Engineering, Innovation Laboratory for Science and Technologies of Energy Materials of Fujian Province (IKKEM) and College of Chemistry and Chemical Engineering Xiamen University, Xiamen 361005, China}}
\affil[4]{\textit{Peng Cheng Laboratory, Shenzhen, China}}
\affil[5]{\textit{School of Materials Science and Engineering, Peking University, Beijing, China}}

\affil[*]{These authors contributed equally to this work}
\affil[$\dagger$]{Corresponding authors: yuanli-ece@pku.edu.cn, xcwang@xmu.edu.cn, fmo@pku.edu.cn
}

\subfile{Content/1abstract}
\begin{document}
\flushbottom
\maketitle
\subfile{Content/2introduction}

\subfile{Content/3results}

\subfile{Content/4discussion}

\subfile{Content/5methods}
\subfile{Content/6DataAvailable}

\subfile{Content/7CodeAvailable}

\bibliography{related_papers}

% \section*{Acknowledgements (not compulsory)}

% Acknowledgments should be brief, and should not include thanks to anonymous referees and editors, or effusive comments. Grant or contribution numbers may be acknowledged.

\section*{Author contributions statement}

% Must include all authors, identified by initials, for example:
% A.A. conceived the experiment(s),  A.A. and B.A. conducted the experiment(s), C.A. and D.A. analyzed the results.  All authors reviewed the manuscript. 

% Yu Wang1,2, Chao Pang1,2, Yuzhe Wang1,2, Junru Jin1,2, Jingjie Zhang1,2, Xiangxiang Zeng3, Ran Su4, Quan Zou 5*, and Leyi Wei
% Y.W., C.P., L.W. conceived the basic idea and designed the research study.
% Y.W. developed the method. 
% Y.W., C.P., J.Z., Y.Z. W. and J.J. evaluated the performance on single-step retrosynthesis. Y.W. devised the explainable decision process. 
% Y.W., C.P., Y.Z. W. searched the literature for multi-step results.

% H.L., D.L., L.Y., Y.H. T., X.C. W. and F.Y. M.
H.L., D.L., X.W., and F.M. conceived the basic idea and designed the research study.
D.L. and F.M. generated the ECD spectra dataset using DFT calculation.
H.L. developed the method. L.Y. and F.M. further modified the method.
D.L. and X.W. conceived the evaluation metric in the experiment.
H.L. and D.L. evaluated the performance on the CMCDS dataset and natural product molecules.
H.L. and D.L. wrote the manuscript. L.Y., Y.T., X.W., and F.M. revised the manuscript.
Y.L. and Y.T. provided the deep learning computing platform.

\clearpage

\section*{Support information}
\subsection*{S1 Statistical results for the CMCDS dataset} 
\begin{figure*}
    %\centering
    \includegraphics[width=1.0\linewidth]{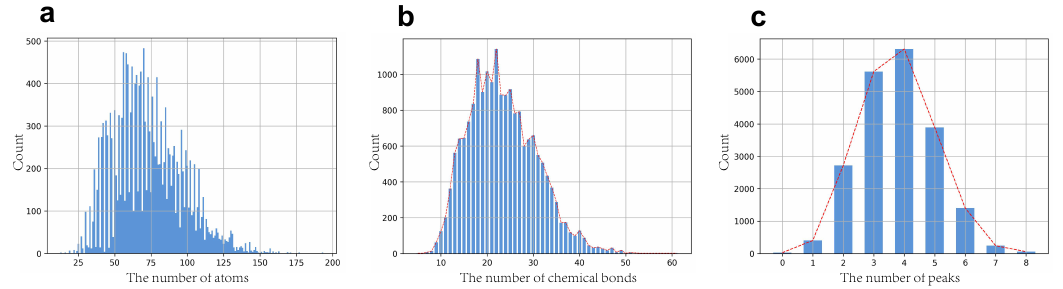}
    \caption{
    \textbf{The visualization of the molecular properties in the CMCDS dataset.} 
    \textbf{a} All molecules contain fewer than 200 atoms, and most molecules have about 75 atoms. \textbf{b} All molecules contain fewer than 65 chemical bonds, and most molecules have about 25 chemical bonds. \textbf{c} The number of peaks ranges from 0 to 8 in the ECD spectrum, and most have 4 peaks.}
    \label{fig:SI_1}
\end{figure*}

%由于我们的数据集中的手性分子都是单手性的，所以它们的复杂度是有限的。可以观察到大多数分子由60个原子左右构成，最大的分子也不会超过200个原子（图a）；同时，它们大多数具备25根化学键左右，最多不超过65根（图b）；而这些分子的ECD谱图大都是具备3/4个峰，最多也不会超过8个峰（图c）。
For all molecules in the CMCDS dataset, we visualized their property distribution by counting the number of atoms in each molecule, the number of peaks of the corresponding ECD spectra, and the number of chemical bonds. In the CMCDS dataset, all chiral molecules are single-chiral-centered, resulting in finite complexity. It is observable that the majority of these molecules consist of approximately 60 atoms, with the largest molecule not exceeding 200 atoms (Fig.~\ref{fig:SI_1}(a)). Furthermore, most of these molecules possess around 25 chemical bonds, with a maximum of 65 bonds (Fig.~\ref{fig:SI_1}(b)). Additionally, the ECD spectra of these molecules typically exhibit 3 to 4 peaks, with a maximum of 8 peaks (Fig.~\ref{fig:SI_1}(c)).

%\subsection*{S2 Comparing the Boltzmann-averaged spectrum with the spectrum of a single conformation.} 
%Before conducting large-scale theoretical calculations, we compared the impact of conformational searching on ECD spectra. Based on experimental results, we observed that the Boltzmann-averaged spectra exhibit a similar overall shape to the spectra obtained from individual conformations, with variations primarily in the peak heights and minor peak shifts. In general, neglecting conformational searching does not significantly affect our final determination of the molecule's absolute configuration. Consequently, when calculating ECD spectra, we will not consider the Boltzmann-averaged spectra.

\newcommand{\pub}[1]{\color{gray}{\tiny{#1}}}
\newcommand{\Frst}[1]{{\textbf{#1}}}
\newcommand{\Scnd}[1]{{\underline{#1}}}
\begin{table*}[]
\centering
% \footnotesize
\renewcommand{\arraystretch}{1.35}  % 控制行高
\setlength{\tabcolsep}{2.0mm}        % 控制列间距
{
{
\begin{tabular}{l|p{80pt}|c|ccc}
    \toprule[1.5pt]
    \textbf{\#} & \textbf{Method} & \textbf{Molecule-Type} & {Position-RMSE~(\textit{nm})~$\downarrow$} & {Number-RMSE~$\downarrow$} & {Symbol-Acc.~(\%)~$\uparrow$} \\
    % \multirow{2}{*}{\textbf{\#}} & \multirow{2}{*}{\textbf{Method}} & \multicolumn{2}{c|}{\textbf{Initialization}} & \multicolumn{3}{c}{\textbf{Evaluation Metrics}}\\ 
    
    % \cmidrule(rl){3-4}\cmidrule(rl){5-7}
    % & & {Rand} & {Pretrain} & {Position-RMSE~(\textit{nm})~$\downarrow$} & {Number-RMSE~$\downarrow$} & {Symbol-Acc.~(\%)~$\uparrow$}\\
    
    \noalign{\hrule height 1.5pt}
    1& ECDFormer & Single-Chiral-Center & 2.29 & 1.24 & 72.7\\ 
    2& ECDFormer & \textbf{Multi}-Chiral-Center & \textbf{2.88} & \textbf{1.76} & \textbf{63.1}\\
 \bottomrule[1.5pt]
\end{tabular}
}
}
% \vspace{-1pt}
\caption{\textbf{Performance for ECD prediction for multi-chiral-centered molecules.}
    We also propose a comparison between the performance of single-chiral-centered molecules and multi-chiral-centered molecules. ECDFormer suffers a slight performance decrease when predicting multi-chiral-centered molecules, demonstrating the generalization ability of ECDFormer.
}
\label{tab:multi-carbon-results}
% \vspace{-1pt}
\end{table*}

\subsection*{S2 The Generalization Ability on Multi-Chiral-Centered Molecules}
Our ECDFormer is trained on the CMCDS dataset, where all molecules have a single-chiral-centered carbon. In Table.~\ref{tab:main_results}, our ECDFormer achieves outstanding ECD spectra prediction performance for single-chiral-centered molecules. To evaluate the generalization ability of our ECDFormer, we further test ECDFormer's performance on multi-chiral-centered molecules, which are more complex in molecular structure and ECD spectra. Specifically, we gather a small group of multi-chiral-centered molecules with their ECD spectra as our test dataset. Then we evaluate the ECDFormer's performance on this multi-chiral-centered dataset. As shown in Table.~\ref{tab:multi-carbon-results}, ECDFormer suffers a slight performance decrease when predicting multi-chiral-carbon molecules, demonstrating the generalization ability of ECDFormer. The good performance of ECDFormer on both single-chiral-centered and multi-chiral-centered molecules further validates its strong applicability.

\subsection*{S3 The Chemical Interpretability Analysis of ECDFormer}

To comprehensively assess the performance of the entire deep learning model, we focused on analyzing the chemical interpretability of the ECDFormer model. By visualizing all the predicted cases in the test split of CMCDS, We found that the spectral similarity of each conformation within a molecule can impact the predictive results of the ECDFormer.

\begin{figure*}
    %\centering
    \includegraphics[width=1.0\linewidth]{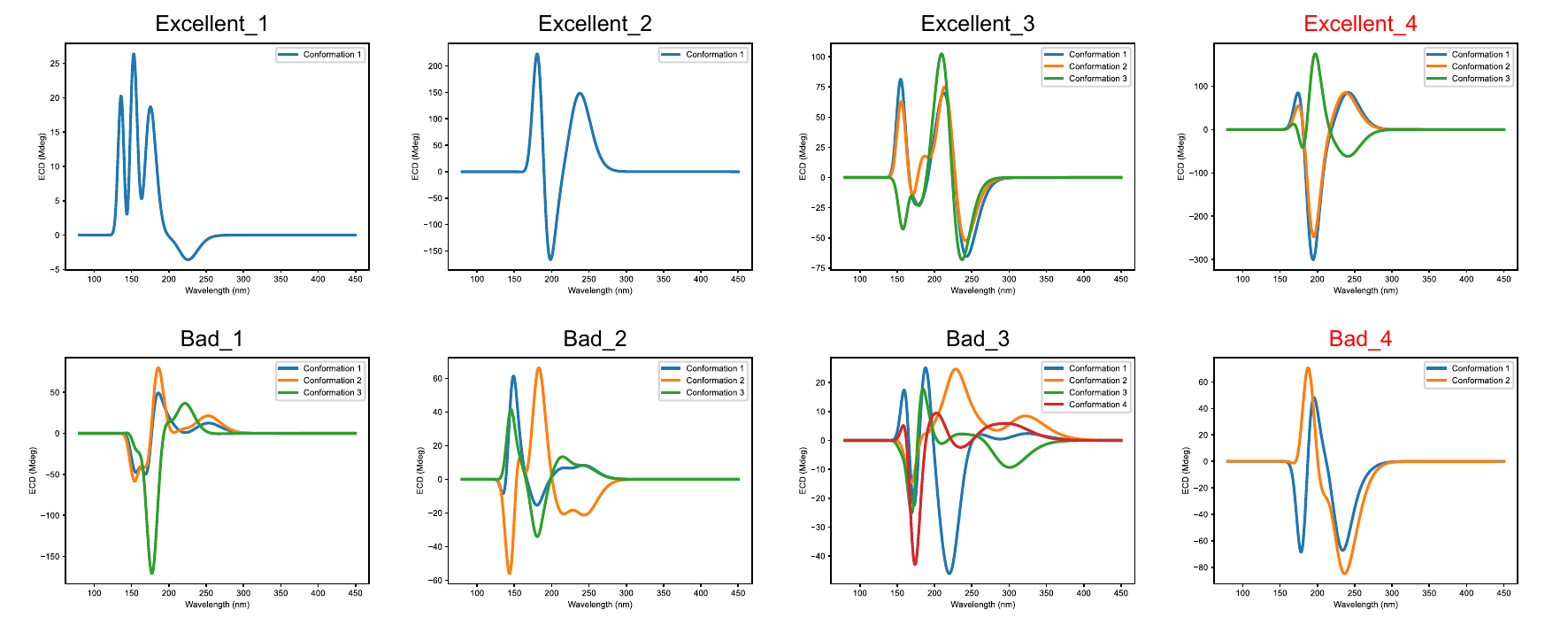}
    \caption{The first column is the calculated ECD spectrum of molecules in the excellent class, obtained after a conformational search with different conformations using the same calculation method. In the second column are the calculated ECD spectra of molecules in the bad class, obtained in the same way as in the first column, for different conformations.}
    \label{fig:SI_CONFORMATION.pdf}
\end{figure*}

Specifically, we selected molecules that were perfect matches and those that were completely wrong, ran conformational searches~\cite{olboyle2011confab} on them, and then calculated the ECD spectra for each conformation. We found that the spectral similarity of each conformation within a molecule can impact the predictive results of the model.
As shown in Fig.\ref{fig:SI_CONFORMATION.pdf}, in the excellent-predicted cases, the ECD spectrum of each conformation showed minor differences. In contrast, in most of the bad-predicted cases, the ECD spectra for each conformation showed significant differences in peak shapes and wavelength. 
This suggests that for different configurations of the same molecule, if their ECD spectra are highly similar, the prediction of ECD spectra by the trained model will be accurate, and vice versa. This phenomenon can be explained from the deep-learning aspect. For a molecule with different ECD spectra shapes, deep-learning models are hard to learn the latent features for prediction. In contrast, for a molecule with similar ECD spectra shapes, deep-learning models are easy to learn the latent pattern from ECD spectra, which improves the prediction performance.
However, a small number of molecules~(Excellent\_4 and Bad\_4 in Fig.\ref{fig:SI_CONFORMATION.pdf}) do not fit this pattern due to the uncertainty in the deep-learning method and the complexity of the chiral assignation field. We are currently investigating these exceptional cases further.

% Specifically, We selected molecules from both the excellent-predicted cases and bad-predicted cases and calculated their ECD spectra for different conformations. As shown in Fig.\ref{fig:SI_CONFORMATION.pdf}, the first three plots in the first column show that the spectra of different conformations of molecules in the excellent category are relatively similar, which seems to reduce the prediction difficulty of the model, but the last plot can still be predicted well even if there are inversions of the corresponding peaks between conformations; Similarly, in the first three plots in the second column, the spectra of different conformations of the molecules in the bad category are very different, which seems to increase the prediction difficulty of the model, but the last plot can still be predicted well even if there is no big difference between conformations. Similarly, in the first three plots of the second column, the spectra of the different conformations of the bad class of molecules are very different, which seems to increase the difficulty of the model prediction. Therefore, we still cannot give a universal explanation from the point of view of molecular conformation, but this also shows the complexity of the chiral field, which is worth exploring further.

\end{document}

%% file: Content/1abstract.tex
\begin{abstract}

% Determining the absolute configuration of chiral molecules holds significant importance in the fields of asymmetric catalysis, functional materials, and the drug industry. The conventional approach requires theoretical calculation of electronic circular dichroism~(ECD) spectra, which is time-consuming and costly. To speed up this process, we have incorporated deep learning techniques for the ECD prediction.
% We first set up a large-scale dataset of Chiral Molecular ECD spetra~(CMCDS) with calculated ECD spectra. We further develop the ECDFormer model, a Transformer-based model to learn the chiral molecular representations and predict their corresponding ECD spectra with improved efficiency and accuracy. 
% Unlike other models for spectrum prediction, our ECDFormer creatively focused on peak properties rather than the whole spectrum sequence for prediction, inspired by the scenario of chiral molecule assignation. Specifically, we applied empty peak property tokens and chiral molecular feature tokens into an encoder to learn the peak properties, including number, position, and symbol. Then we render the ECD spectra from these peak properties, which significantly outperforms other models in ECD prediction, Our ECDFormer reduces the time of acquiring ECD spectra from a few hours or dozens of hours per molecule to 10ms.

Chiral molecule assignation is crucial for asymmetric catalysis, functional materials, and the drug industry. The conventional approach requires theoretical calculations of electronic circular dichroism~(ECD) spectra, which is time-consuming and costly. To speed up this process, we have incorporated deep learning techniques for the ECD prediction. We first set up a large-scale dataset of Chiral Molecular ECD spectra~(CMCDS) with calculated ECD spectra. We further develop the ECDFormer model, a Transformer-based model to learn the chiral molecular representations and predict corresponding ECD spectra with improved efficiency and accuracy. Unlike other models for spectrum prediction, our ECDFormer creatively focused on peak properties rather than the whole spectrum sequence for prediction, inspired by the scenario of chiral molecule assignation. Specifically, ECDFormer predicts the peak properties, including number, position, and symbol, then renders the ECD spectra from these peak properties, which significantly outperforms other models in ECD prediction, Our ECDFormer reduces the time of acquiring ECD spectra from 1-100 hours per molecule to 1.5s.

\textbf{Keywords}

Chiral Molecule Assignation, ECD Spectra Prediction, Deep Learning.

% 由于其化学性质的截然不同，手性分子对的区分问题在化学分子合成/制药领域非常重要.
% 然而传统的(手性分子对)区分技术需要理论计算出圆二色光谱, which is extremely time-consuming.
% 在本论文中我们首次提出使用机器学习的方法来解决理论计算圆二色光谱的问题
% 我们首先构建了一个大规模手性分子与光谱对应的数据集
% 基于该数据集, 我们提出一种xxx 模型, 能够
% 实验证明, 我们的模型可以准确的预测给定手性分子的光谱结构, 

\end{abstract}

%% file: Content/2introduction.tex
% \section{Introduction}

\begin{figure*}
    %\centering
    \includegraphics[width=1.0\linewidth]{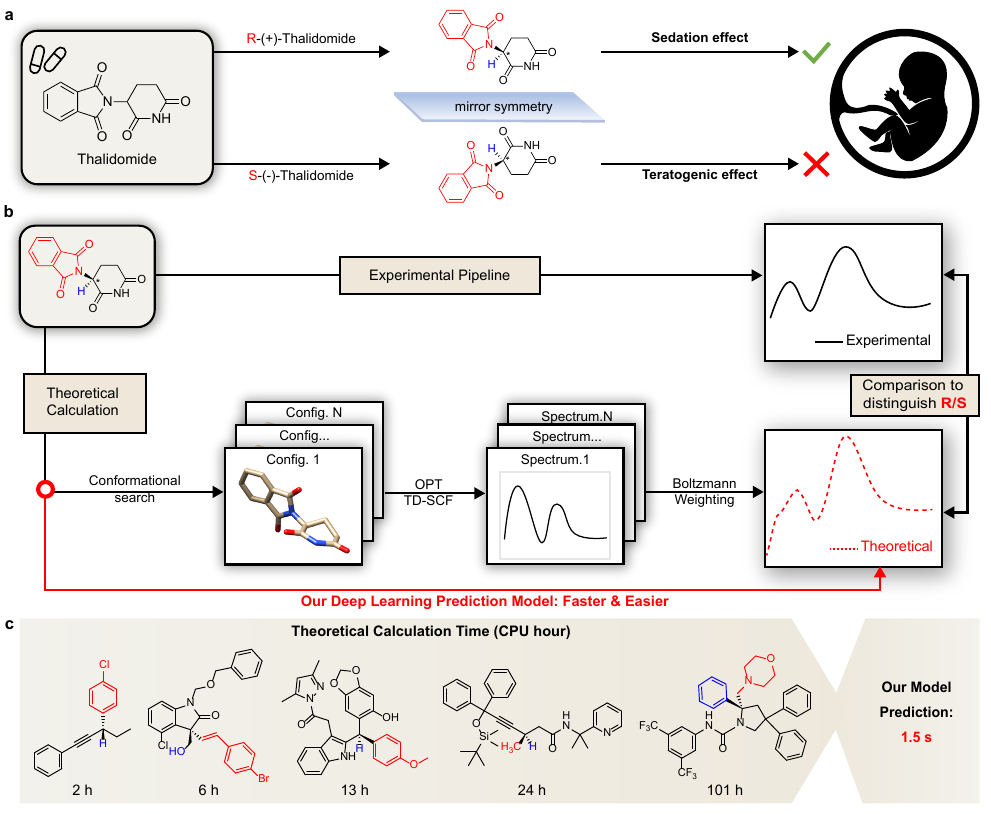}
    \caption{
    \textbf{The scheme for ECD prediction and chiral molecule assignation.} 
    \textbf{a} Thalidomide has two configurations (R/S). R-Thalidomide induces sedative effects, whereas S-Thalidomide is associated with teratogenic effects. 
    \textbf{b} ECD comparison is most frequently employed for assigning the absolute configuration. However, The theoretical calculation of ECD is time-consuming, involving steps such as conformational searching, conformational optimization, excited-state property calculation, and Boltzmann weighting. So we employ deep learning for acceleration. 
    \textbf{c} As molecules become more complex, the computation time increases. Our CPU version is \textit{IntelXeonE5-2640v4@2.40GHz}.
    }
    \label{fig:intro_fig}
\end{figure*}

A chiral molecule refers to a unique spatial arrangement that cannot be superimposed onto its mirror image, resulting in non-identical left-handed and right-handed forms. Chirality is ubiquitous in chemistry and biology and plays a crucial role in various fields such as asymmetric catalysis~\cite{noyori2002asymmetric,list20212021}, functional materials~\cite{amabilino2006supramolecular,shen2020supramolecular}, drug discovery~\cite{teng2022advances}, and other related areas~\cite{lininger2023chirality,evers2022theory}. Specifically, in the drug discovery area, the drug activity often depends on its absolute configuration. A well-known chiral drug is thalidomide in Fig.~\ref{fig:intro_fig}(a), which was previously used as an antiemetic drug for morning sickness~\cite{zhang2019great} in the form of enantiomeric pairs. However, one of its chiral configurations~(R-type) is safe, while the other chiral configuration~(S-type) induces severe teratogenic effects. Thus, assigning the absolute configuration of chiral molecules has always been the center of chiral-related research.

There are traditional approaches for discerning the chiral configuration of a molecule with single chiral carbon, including electronic circular dichroism~(ECD) spectroscopy, nuclear magnetic resonance spectroscopy, and X-ray single-crystal diffraction methods~\cite{ebeling2018assigning,menna2019challenges}.
Among these methods, ECD spectroscopy is the most efficient and reliable method for determining the absolute configuration of chiral molecules. However, the procedure is still laborious and time-consuming, including chiral separation of isomers, obtaining experimental CD spectra, computation of the theoretical ECD spectra through quantum chemical calculations, and comparison of both experimental and theoretical ECD spectra to achieve conclusive identification of the absolute configuration. Specifically, this comparison focused on the wavelength, the signs of the Cotton effects~(positive or negative peaks), the intensity of peaks, and their agreement between experimental and calculated spectra~\cite{junior2023absolute}. 

For experimental chemists, the theoretical calculation of ECD spectra in the aforementioned steps stands out as the most time-consuming and technically demanding task. As shown in Fig.~\ref{fig:intro_fig}(b), the computation of ECD spectra for a chiral molecule entails multiple stages. Initially, a molecular structure model is drawn, followed by molecular dynamics simulations to explore various energetically favorable conformations. Subsequently, these conformations undergo individual structure optimization and energy calculations at the density functional theory~(DFT) level of precision. Then the ECD spectra of the molecules are computed employing time-dependent DFT~(TD-DFT) calculations. The final calculated ECD spectrum is generated by combining the individual ECD spectra of different conformations, weighted by their Boltzmann probabilities. This requires experimental chemists to possess a proficient understanding of specialized tools, such as molecular dynamics and DFT calculations. Moreover, the computational demands and time requirements associated with this process are substantial, thereby highlighting its rate-determine step in the assignment of chiral absolute configurations. It raises an open question: ``\textit{Can we speed up the theoretical calculation of ECD spectra?}''

% Recently, the rapid development of deep learning has brought prosperity to the field of chemistry~\cite{xu2022high,xu2023retention} due to its efficiency. Thus, we are the first to model the ECD theoretical calculation process as a sequence prediction task and use deep-learning models to speed up the process. 
% Datasets are fundamental to deep learning models since the quantity and quality of the dataset are directly related to the performance of the deep learning models.
% Therefore, we propose an effective auto-collecting system to extract the chiral molecular pairs and their corresponding ECD spectra from related datasets. Specifically, shown in Fig.~\ref{fig:dataset}, we clean and select the chiral molecular pairs from the chiral molecular dataset: CMRT~\cite{xu2023retention}. Then we used Gaussian software~\cite{g16} to calculate the ECD spectra of each chiral molecule, which takes 5,0000 CPU hours. Finally, we combine the 22190 chiral molecules and their ECD spectra as the \textbf{C}hiral \textbf{M}olecular \textbf{CD} \textbf{S}pectrum Dataset~(CMCDS).

In recent years, statistical tools based on machine learning have been integrated into chemistry research workflows~\cite{janet2020machine}. This integration is enabling researchers to analyze vast datasets with greater precision and discover intricate patterns and relationships that were previously undetectable, significantly enhancing the efficiency and effectiveness of chemical research and innovation~\cite{de2019synthetic,hermann2022ab}. Large and high-quality datasets are essential for the effectiveness of machine learning methods. We first need to have a library of chiral molecules. Fortunately, we have constructed a library of 25000+ chiral molecules~(Chiral Molecules Retention Time Dataset, CMRT) in our previous work, which introduced a machine learning framework to enhance the efficiency of chromatographic enantioseparation in experimental chemistry~\cite{xu2023retention}. Based on the CMRT dataset, a Chiral Molecular CD Spectra Dataset~(CMCDS) was generated by selecting chiral molecules from CMRT and calculating their ECD spectra. To the best of our knowledge, CMCDS is the first large-scale dataset for ECD spectra prediction.

With the CMCDS dataset, we further construct the ECDFormer, a deep-learning model to speed up the prediction of the ECD spectra for chiral molecules. Inspired by the chemical assignation scenario that focuses on peak properties in the ECD spectra, our ECDFormer creatively proposes a peak property prediction module to render the ECD spectra from peak properties rather than predict the ECD spectra directly.
% ~\cite{zou2023deep,shen2019molecular}. 
For the input molecule, our ECDFormer applies its atom, bond, angle features, and molecular descriptors as the description information into the GeoGNN structure~\cite{fang2022geometry} to learn the molecular representation. For the peak property learning module, we apply the transformer encoder~\cite{vaswani2017attention} to learn the peak property features from molecular representations. Then we respectively predict the peak number, position~(wavelength), and symbol~(the sign of Cotton effect) from property features and render them into the ECD spectra as the prediction of theoretical ECD spectra. 

The quantitative experimental results demonstrate the accuracy and efficiency of our ECDFormer compared with other baselines that directly predict the whole ECD spectra. The visualizations show that ECDFormer predicts correct ECD spectra for molecules in CMCDS as well as the natural molecules with pharmaceutical effects. Our model not only advances research in chiral chemistry but also has potential applications in asymmetric synthesis and facilitates high-throughput screening of chiral drug molecules in the pharmaceutical development field. Our contribution can be summarized as follows:
\begin{itemize}
\setlength{\itemsep}{0pt} % 缩小行间距
    % \item The ECD spectra calculation for chiral molecular assignation is crucial yet time-consuming for chemists. We propose a deep learning model, ECDFormer, to predict the ECD spectra efficiently. We also propose the CMCDS dataset for the ECD prediction task, containing the ECD spectra and SMILES of 22,190 chiral molecules.

    % \item Inspired by the procedures of determining the absolute configuration in chemistry, our ECDFormer focuses on the peak prediction and renders peaks into the ECD spectra, instead of predicting the whole spectra sequence like other baselines. The peak prediction module in ECDFormer significantly improves the performance of ECD prediction.

    % \item Experimental results show that our model can accurately predict the ECD spectra using SMILES of the chiral molecule, and thus speed up the chiral configuration assignation process.

    \item The ECD spectra calculation for chiral molecular assignation is crucial yet time-consuming for chemists. A deep-learning model, ECDFormer, was proposed to predict the ECD spectra and improve the assignation efficiency. Inspired by the assignation procedure in chemistry, ECDFormer focuses on peak prediction and renders peaks into the ECD spectra.
    
    \item We proposed a large-scale dataset, CMCDS, for the ECD prediction task. CMCDS containing ECD spectra for 22,190 chiral molecules was produced utilizing substantial computational power.

    \item Experimental results demonstrate the accuracy and efficiency of ECDFormer on the CMCDS dataset. ECDFormer also predicts correct ECD spectra for the natural product molecules that have pharmaceutical effects.
    
\end{itemize}

%% file: Content/3results.tex
\section{Results}

\subsection{Construction of the CMCDS dataset} 

% Molecular chirality, as a pervasive phenomenon in nature, is an important factor influencing the properties of molecules. Chiral molecules usually exist in pairs, often referred to as enantiomers, which are mirror images of each other but are unique. Currently, there are three main methods for distinguishing enantiomers, based on NMR and X-ray diffraction, which are very time-consuming and inaccurate, and based on actual circular dichroism, combined with theoretically calculated electronic circular dichroism, which is the more conventional method for distinguishing enantiomers. However, to obtain the circular dichroism of a molecule, it is necessary to obtain multiple conformations by conformational search, optimize these conformations, further calculate the circular dichroism of each conformation, and finally obtain the circular dichroism by Boltzmann averaging. The whole computational process is still complex and time-consuming, and this work mainly aims to accelerate the acquisition of electronic circular dichroism of molecules through deep learning methods, thus accelerating the identification of chiral molecules.

\begin{figure*}[h]
    % \centering
    \includegraphics[width=1.0\linewidth]{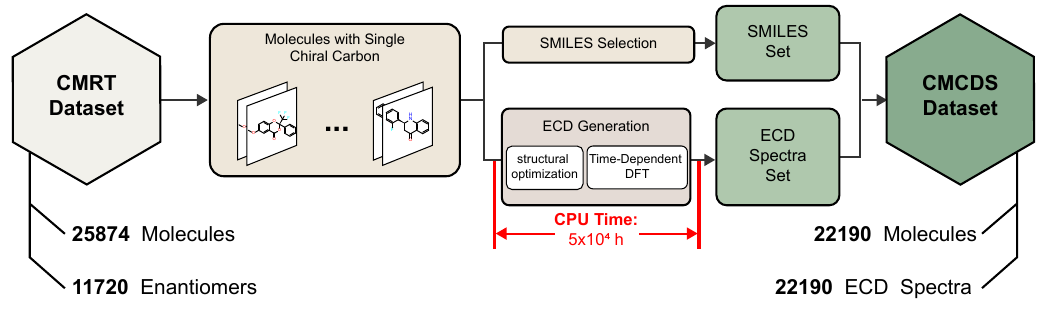}
    \vspace{-0.05in}
    \caption{\textbf{The generation pipeline} for our chiral molecular CD spectra dataset~(CMCDS) for ECD prediction task. }
    \label{fig:dataset}
    \vspace{-0.1in}
\end{figure*}

As shown in Fig.\ref{fig:dataset}, the CMCDS dataset is mainly realized by large-scale theoretical calculations, consisting of ECD spectra and SMILES sequences of 22190 chiral molecules, and the ECD spectral data of all the molecules were calculated by Gaussian16
A.03 packages~\cite{g16}. Our chiral molecules were mainly crawled from the literature of asymmetric catalysis, and we transformed the SMILES files of the molecules into MOL files with the help of the RDKit package to obtain the 3D atomic coordinates of the molecules. The above MOL files were converted into Gaussian input gjf files in batches through Python. Then the molecule structure was optimized at B3LYP~\cite{stephens1994ab}/6-31G level.  Furthermore, we conducted the electronic circular dichroism calculation at the CAM-B3LYP~\cite{yanai2004new}/6-31G(d) level, setting the number of states~(nstates) to 20. We fix the half-peak width at 0.3 and apply Gaussian broadening, utilizing the energies and wavelengths derived from these 20 excited states. The ECD spectra of all molecules were acquired in the same way, and we used Python for batch data processing.

% To achieve this, the CMCDS dataset is mainly realized by large-scale theoretical calculations, which will be provided in the Methods section. The dataset consists of ECD spectra and SMILES sequences of 20,000 chiral molecules, providing ample data for the task of predicting ECD spectra.

\subsection{Construction of the ECDFormer model}
\begin{figure*}[h]
    % \centering
    \includegraphics[width=1.0\linewidth]{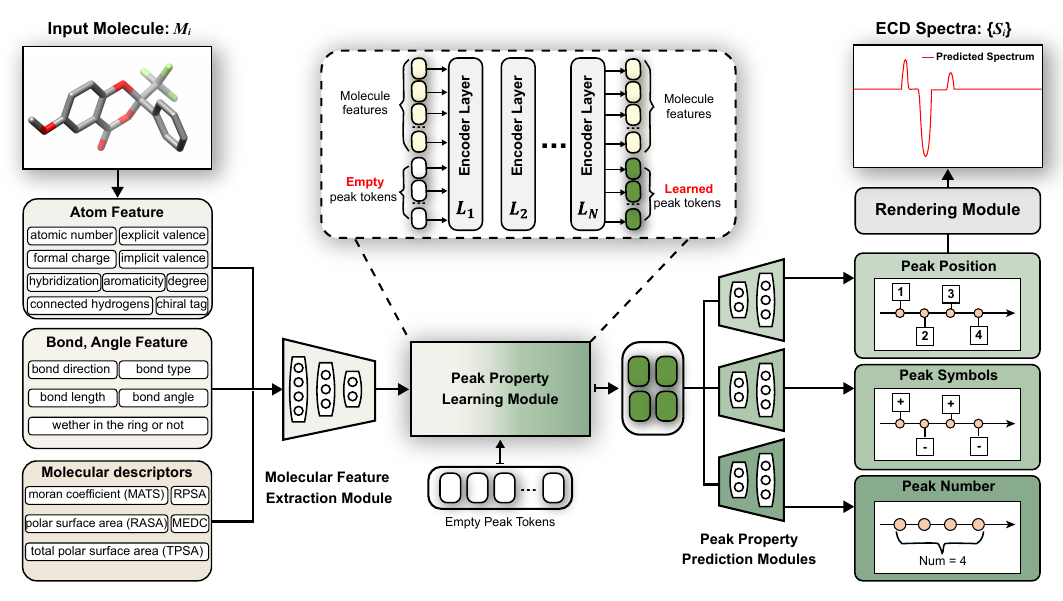}
    \vspace{-0.2in}
    \caption{
    \textbf{The General Pipeline of our ECDFormer model.} The design of the peak property learning and prediction modules is inspired by the chemical chiral assignation procedure. By predicting peak properties and rendering ECD spectra, ECDFormer outperforms baselines in the ECD spectra prediction task. 
    }
    \label{fig:framework}
    \vspace{-0.3in}
\end{figure*}

Fig.\ref{fig:framework} shows the computational workflow of our ECDFormer model. The workflow takes the atom-bond-angle features and molecular descriptors as the features of the target molecule.  
ECDFormer contains four modules for ECD prediction: 
\textbf{(i)}~the molecular feature extraction module to get the chiral molecular representation based on a geometric-enhanced graph neural network.
\textbf{(ii)}~the peak property learning module to extract the peak property features from chiral molecular representation using a Transformer Encoder structure. 
\textbf{(iii)}~the peak property prediction module to predict the peak properties, including number, position, and symbol, from the learned peak property features. 
\textbf{(iv)}~the ECD rendering module to reconstruct the ECD spectra from predicted peak properties.

% Different molecules, owing to their distinct structures, exhibit variations in peak positions and peak intensities in spectra. In ECD spectra, compared with other types of spectroscopic techniques, the most prominent feature distinguishing spectra between different molecules, aside from peak positions and intensities, is the occurrence of positive and negative peaks due to the Cotton effect~\cite{zou2023deep}. Thus, traditional sequence prediction models perform badly on the ECD prediction task. Chemists focused on the peaks’ symbols~(corresponding to the sign of the Cotton effect) and positions~(corresponding to the wavelength of the peaks) in ECD spectra for chiral molecular assignation. Therefore, to simplify the ECD prediction task, we predict the key information in ECD, including peak numbers, positions, and symbols. We also design a peak-focused loss function accordingly.

Molecular Electronic Circular Dichroism~(ECD) spectra are characterized by the presence of positive and negative peaks as a result of the Cotton effect~\cite{zou2023deep}. 
Compared to other spectra including protein ECD spectra~\cite{rogers2019electronic} and molecular infrared spectra~\cite{zou2023deep}, molecular ECD spectra reveal significant morphological variations. This distinct feature makes traditional sequence prediction models~(LSTM, GRU) less effective for ECD prediction by directly predicting the whole spectra. 
Chemists often concentrate on the symbols of peaks (indicating the direction of the Cotton effect) and their positions (related to the wavelengths of the peaks) in ECD spectra for determining chirality in molecules. To streamline the ECD prediction process, we focus on predicting essential ECD information such as the number of peaks, their positions, and symbols. Accordingly, the peak-focused loss function to support this approach is:

\begin{gather}
    L(y^{true}, y^{pred}) = L_{ce}^{Num}(y^{true}, y^{pred}) + (L_{ce}^{Pos}(y^{true}, y^{pred}) + 2*L_{ce}^{Sym}(y^{true}, y^{pred}))
\end{gather}
where $L_{ce}$ for peak number, position, and symbol are cross-entropy loss~\cite{mao2023crossentropy}. Due to the emphasis of ECD spectra prediction on the positive and negative peaks, we slightly increased the loss weight for peak symbols to enforce the model prediction.

\subsection{Peak-specific Evaluation Metrics for the ECD Prediction Task}
% by Lihao 11.2
% 化学分子的ecd光谱存在两个特点 (i).形状丰富度很大 (ii).在指认手性分子时很依赖序列的峰属性，which和蛋白质的ecd光谱截然不同，因此我们无法将沿用蛋白质ecd光谱预测任务~\cite{nagy2019sesca}中的RMSE评测指标。为了更好评测ecd光谱预测任务以及后续的分子指认中模型的性能，我们围绕ecd光谱的峰属性建立了三套评测指标来从不同角度合理的评测ecd光谱预测质量

The ECD spectra of chemical molecules exhibit two distinct characteristics: (i). a high degree of shape diversity, (ii). a strong reliance on peak attributes for chiral molecule identification. These characteristics are significantly different from the ECD spectra of proteins, rendering it inappropriate to adopt the Root Mean Square Error~(RMSE) evaluation metric used in protein ECD spectrum prediction tasks~\cite{nagy2019sesca,micsonai2021bestsel,zhao2021accurate}. 
To better evaluate the quality of the ECD spectrum for the chiral molecular assignation task, we establish three sets of evaluation metrics based on peak attributes of ECD spectra: (1). \textbf{Number-RMSE}: the RMSE of peak number between ground-truth and prediction ECD spectra. (2). \textbf{Position-RMSE}: the RMSE of each peak's position between ground-truth and prediction ECD spectra. (3). \textbf{Symbol-Acc}: the matching accuracy of peaks' symbols between ground-truth and prediction ECD spectra. These metrics provide a reasonable and comprehensive assessment of ECD spectrum prediction quality from different perspectives.

\subsection{Performance comparison on the CMCDS dataset}
% by lihao, 为了更全面的评测我们的ChirlFormer性能, 我们实现了三类模型作为我们的baseline, 包括机器学习模型, 预训练深度学习模型, 以及非预训练深度学习模型. Table.~\ref{tab:main_results}显示我们模型在这三类baseline中取得了SOTA性能. 具体的实验分析如下

To comprehensively evaluate the performance of our ECDFormer, we implemented two categories of models as our baselines, the machine learning models and deep learning models. Table.~\ref{tab:main_results} demonstrates that our model achieves state-of-the-art performance across these baselines. The specific experimental analysis is provided below.

% 以regressor和classifier为代表的机器学习模型是 化学与材料学领域常见的分析工具~\cite{}. 由于手性分子的光谱预测为回归任务, 因此我们也评测了多个回归结构的机器学习模型的性能, 包括SGD-Regressor, PositiveAggressive-Regressor, Logistic-Regressor. 表格1展示了机器学习模型的性能. 模型在RMSE-Number, Position, Range三个指标上都不理想, 分别比深度学习模型平均低了 XXXX, 这是因为模型需要从复杂分子结构特征中解耦出光谱序列, 当前机器学习模型普遍无法处理复杂分子结构特征, 也无法实现对复杂序列结构的建模, 这也说明了采用深度学习模型解决手性分子光谱预测任务的必要性

\subsubsection{Comparison with machine learning baselines. }
Machine learning models are commonly used as analytical tools in the fields of chemistry and materials science~\cite{artrith2021best,wei2019machine}. We select three common models, including SGD Regressor, Positive Aggressive Regressor, and Logistic Regressor, as the baselines. Comparing line.1-3 and line.10 in Table.\ref{tab:main_results}, machine learning baselines perform unsatisfactorily, which is mainly attributed to the models' inability to decouple spectral sequences from complex molecular structural features. This emphasizes the necessity of employing deep learning models to tackle the task of predicting ECD spectra for chiral molecules.

% Since the prediction of spectra for chiral molecules is a regression task, we also evaluated the performance of multiple regression-based machine learning models, including SGD Regressor, Positive Aggressive Regressor, and Logistic Regressor. Line~1-3 in Table.\ref{tab:main_results} present the performance of these machine learning models. However, the models perform unsatisfactorily, exhibiting an average decrease compared to deep learning models. This is mainly attributed to the models' inability to decouple spectral sequences from complex molecular structural features. Machine learning models struggle with handling complex molecular structural features and modeling complex sequence structures. This emphasizes the necessity of employing deep learning models to tackle the task of predicting spectra for chiral molecules.

% 基于分子表征学习的预训练模型~\cite{}近年来被充分探索. 由于在大规模分子数据集上训练, 分子表征预训练模型能在常见下游任务, 比如分子结构预测, 分子性质预测上去的良好效果. 然而分子表征预训练模型无法有效解决手性分子相关问题. line~4-5展示了两个分子表征预训练模型的性能, 尽管在RMSE-Number和RMSE-Position上有提升, 但性能依然与我们的ChirlFormer有差距, 这是因为分子表征预训练模型的预训练任务无法区分手性这一细粒度结构差异,因此无法对手性分子进行可区分的有效建模
\subsubsection{Comparison with deep learning baselines. }
In the context of abundant data, deep learning models have shown excellent performance in complex task settings. With the CMCDS dataset, we implement sequence prediction deep learning models as our baselines, including LSTM~\cite{shi2015convolutional}, GRU~\cite{chung2014empirical}, and Transformer Decoder~\cite{vaswani2017attention}. Comparing line.5-8 with line.10 in Table.\ref{tab:main_results}, our ECDFormer, predicting the peak property of ECD spectra, significantly outperforms other baselines. The results demonstrate the effectiveness of our peak property prediction module in ECDFormer. Comparing line.6/8 with line.7/9, the pretrained models have little influence on the ECD prediction task, due to the lack of chiral molecular information during the pretraining stage

% In recent years, pre-trained models based on molecular representation learning~\cite{} have been extensively explored. Trained on large-scale molecular datasets~\cite{}, pre-trained models exhibit promising performance on common downstream tasks such as molecular structure prediction~\cite{} and molecular property prediction~\cite{}. However, they are not effective in addressing chiral molecule-related problems. Line.4-5 present the performance of two pre-trained models for molecular representation learning. Although there is improvement in RMSE-Number and RMSE-Position, there still exists a performance gap compared to our ChirlFormer in line.11. This is because the pre-training tasks of molecular representation models fail to differentiate the fine-grained structural differences associated with chirality, making it challenging to effectively model chiral molecules in a distinguishable manner.

\newcommand{\pub}[1]{\color{gray}{\tiny{#1}}}
\newcommand{\Frst}[1]{{\textbf{#1}}}
\newcommand{\Scnd}[1]{{\underline{#1}}}
\begin{table*}[]
\centering
\footnotesize
\renewcommand{\arraystretch}{1.35}  % 控制行高
\setlength{\tabcolsep}{4.0mm}        % 控制列间距
{
{
\begin{tabular}{l|p{80pt}|cc|ccc}
    \toprule[1.5pt]
    \multirow{2}{*}{\textbf{\#}} & \multirow{2}{*}{\textbf{Method}} & \multicolumn{2}{c|}{\textbf{Initialization}} & \multicolumn{3}{c}{\textbf{Evaluation Metrics}}\\ 
    
    \cmidrule(rl){3-4}\cmidrule(rl){5-7}
    & & {Rand} & {Pretrain} & {Position-RMSE~(\textit{nm})~$\downarrow$} & {Number-RMSE~$\downarrow$} & {Symbol-Acc.~(\%)~$\uparrow$}\\
    
    \noalign{\hrule height 1.5pt}
    \rowcolor{gray!20}\multicolumn{7}{c}{\it{\textbf{Machine Learning Methods}}} \\
    \hline
    1& Logistic-Regressor & \checkmark & - & 7.81 & 7.22 & 47.8 \\
    2& SGD-Regressor & \checkmark & - & 6.44 & 6.36 & 47.1 \\
    3& Aggr-Regressor & \checkmark & - & 5.97 & 4.39 & 48.5 \\ 
    \noalign{\hrule height 1.0pt}
    \rowcolor{gray!20}\multicolumn{7}{c}{\it{\textbf{Deep Learning Methods}}} \\
    \hline
    4& GeoGNN+Linear & \checkmark & - & 8.62 & 2.87 & 51.9 \\ 
    5& GeoGNN+GRU & \checkmark & - & 6.47 & 1.72 & 39.5 \\
    6& GeoGNN+LSTM & \checkmark & - & 5.91 & 1.76 & 43.7 \\ 
    7& GeoGNN+LSTM & - & \checkmark & 4.68 & 1.45 & 46.4 \\
    8& GeoGNN+Transformer & \checkmark & - & 4.69 & 1.36 & 49.2 \\
    9& GeoGNN+Transformer & - & \checkmark & 5.82 & 1.64 & 37.3 \\
    % \hline
    \noalign{\hrule height 1.0pt}
    \rowcolor{aliceblue!60} 10& \textbf{ECDFormer}~(ours) & \checkmark & - & \textbf{2.29} & \textbf{1.24} & \textbf{72.7}\\
 \bottomrule[1.5pt]
\end{tabular}
}
}
\vspace{-0.05in}
\caption{\textbf{Performance for ECD prediction task.}
    We propose the experimental results on our ECDFormer framework and the corresponding baselines including machine learning models and deep learning models. Focusing on peak property prediction, our ECDFormer model surpasses baselines under all evaluation metrics.}
\label{tab:main_results}
% \vspace{-0.3in}
\end{table*}

\begin{figure*}
    %\centering
    \includegraphics[width=1.0\linewidth]{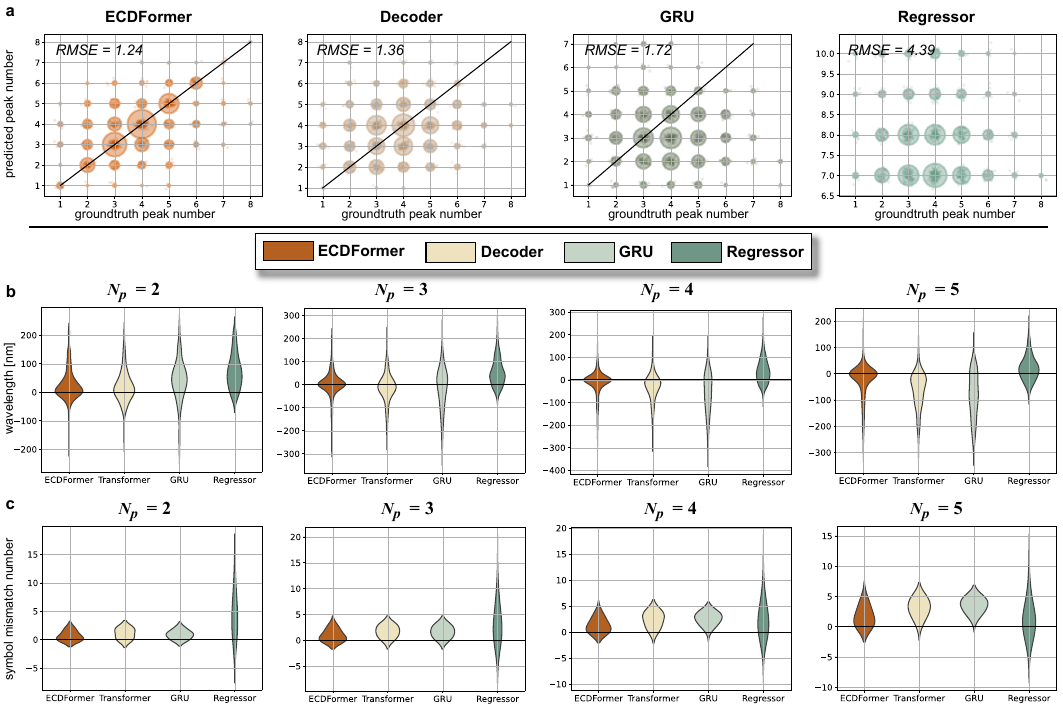}
    \caption{
    \textbf{The performance comparison between ECDFormer and baselines for ECD prediction.} 
    \textbf{a} The data distribution plot for the ground-truth peak number and their predicted number.
    \textbf{b} The violin plot of the discrepancies in peak positions between ground-truth ECD and predicted ECD from ECDFormer and baselines. $N_{v}$ is the peak number, representing the difficulty of cases.
    \textbf{c} The violin plot of the discrepancies in peak symbols between ground-truth ECD and predicted ECD from ECDFormer and baselines.
    }
    % \textbf{a} In comparing the discrepancies in wavelength between the spectral peaks within each ECD spectrum and the corresponding predicted spectral peaks, a consistent observation emerges: the variance in spectral peak positions falls well within the range of ±10, resulting in a final RMSE of 2.08. 
    % \textbf{b} Based on the number of spectral peaks, the samples were categorized into three groups: "easy" with fewer than two peaks, "medium" with 3-4 peaks, and "hard" with 5-8 peaks. When comparing the correctness of positive and negative value predictions for each peak, it is evident that the majority of samples have accurate predictions, resulting in a final accuracy of 0.793. 
    % \textbf{c} When comparing the actual count of peaks with their predicted values, the resulting RMSE is 1.01.
    \label{fig:model_validation}
\end{figure*}

\subsection{The Analysis Visualization on Peak-specific ECD Evaluation Metrics} 
% 为了更好的展示测试集中预测的ecd光谱在三个
To better analyze the models' performance, including our ECDFormer and other baselines, under three peak-specific evaluation metrics, we draw the analysis graphs for each evaluation metric in Fig.~\ref{fig:model_validation}. The detailed analysis is as follows: 

\subsubsection{Peak Number Analysis}
% 在Figure.a中我们进一步分析了模型对峰的数量的预测能力, 并证明了模型在简单和复杂的光谱的峰值预测上都取得很好的效果.
% Fig.a中X轴表示Peak number的Groundtruth值, Y轴表示Peak number的预测值，因此数据点越靠近y=x,预测性能越好。红色圈表示数据点的密集程度, 其中红圈越大表示数据点越多。
In Fig.~\ref{fig:model_validation}(a), we analyze ECDFormer's predictive capability regarding the peak number and demonstrate its excellent performance in predicting peak number for complex spectra~(Peak-Number$>5$) compared to baseline models. The X-axis represents the ground truth values of the peak number, while the Y-axis represents the predicted values of the peak number. Therefore, the closer the data points are to the $y=x$ line, the better the predictive performance. The density of the data points is indicated by the size of the red circles, where a larger red circle represents a higher concentration of data points. Fig.~\ref{fig:model_validation}(a) shows that in ECDFormer, the largest red circles all appear on the $y=x$ line, even when predicting hard samples~(Peak-Number$>5$). The RMSE of peak number is 1.01, indicating the good performance of peak number prediction for our ECDFormer.

\subsubsection{Peak Position Analysis}
In Fig.~\ref{fig:model_validation}(b), we analyze the model's peak position predictive capability. Specifically, we visualize the violin graphs of the position differences between predicted peaks and ground-truth peaks. To further visualize the performance in easy-to-hard cases, we split the test dataset based on the peak number $N_{v}$ of a molecule. Compared with baselines, for all cases from easy to hard, most predictions in ECDFormer have 0 difference with ground truth, demonstrating the effectiveness.

% The X-axis represents the sequential positions of peaks, while the Y-axis represents the difference between predicted positions and ground-truth positions. Therefore, the closer the data points are to the $y=0$ line, the better the predictive performance. The density of the data points is indicated by the size of the red circles, where a larger red circle represents a higher concentration of data points. Fig.~\ref{fig:model_validation}(b) shows that the errors in most of the peak position predictions are close to zero. The RMSE is 2.08, indicating the good prediction performance of our framework.

\subsubsection{Peak Symbol Analysis}
In Fig.~\ref{fig:model_validation}(c), similar to the peak position analysis, we further analyze the model's peak symbol predictive capability. we visualize the violin graphs of the symbol differences between predicted peaks and ground-truth peaks. Compared with baselines, for all cases from easy to hard, most predictions in ECDFormer have the same symbols as ground truth, demonstrating the effectiveness.

% A further comparison of the signs of the peaks at the same positions was made. By categorizing samples based on the number of peaks in the spectra, three classes were defined for comparison. Simple samples have no more than 2 peaks, moderate samples have 3-4 peaks, and challenging samples have 5-8 peaks. All three classes had a higher probability of fewer incorrectly predicted peaks, and for the majority of samples, the signs of the peaks were entirely correct, resulting in a final accuracy of 79.3\%. 

\subsection{The Visualization of ECD Spectra Prediction Cases}
\begin{figure*}
    % \centering
    \includegraphics[width=1.0\linewidth]{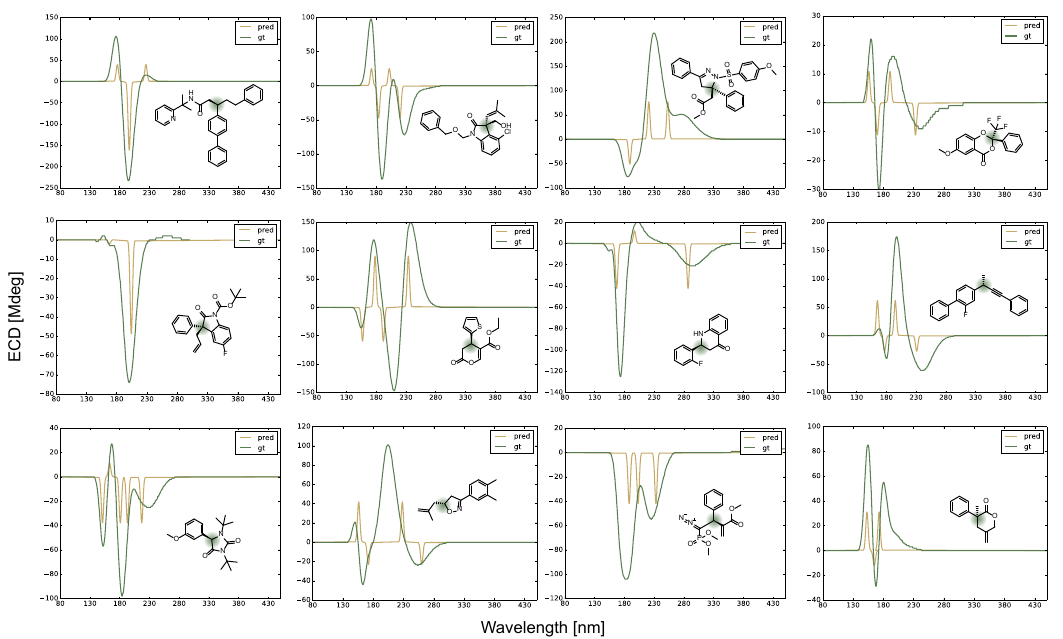}
    \caption{\textbf{Visualization of ECD spectra predictions from ECDFormer}. We visualize the ground-truth spectra and ECDFormer's prediction spectra of the selected molecules from the test split of the CMCDS dataset.}
    \label{fig:visualize}
\end{figure*}

\begin{figure*}
    % \centering
    \includegraphics[width=1.0\linewidth]{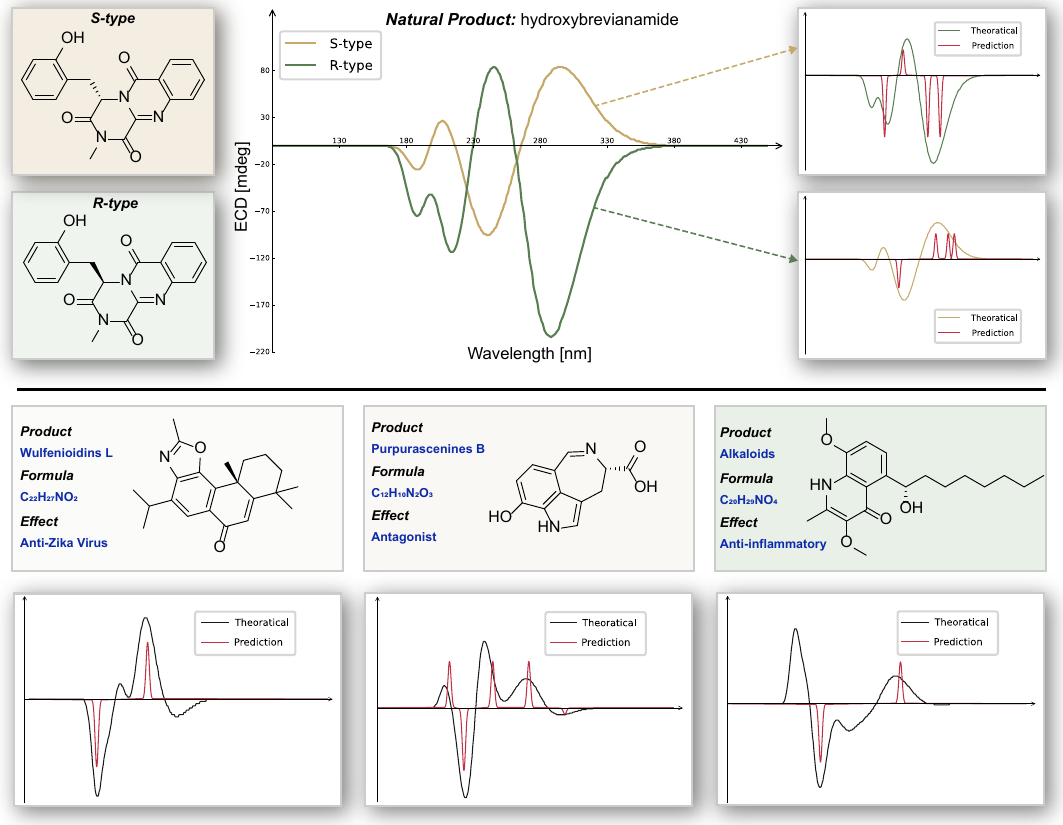}
    \caption{\textbf{ECD predictions on natural products with pharmaceutical effects.} We select pharmaceutical products from recent journals. Visualizations show that ECDFormer can produce correct predictions for natural products and their R/S types.}
    \label{fig:single_drug}
\end{figure*}

Our visualization contains two parts: (a).~Visualizing the ECD spectra corresponding to molecules in the test split of the CMCDS dataset, and (b).~Visualizing the ECD spectra corresponding to existing pharmaceutical molecules. 
Fig.~\ref{fig:visualize} presents our visualization of the CMCDS dataset test split, demonstrating our model's ability to achieve good performance predictions even when faced with complex molecules of various structures. 
Fig.~\ref{fig:single_drug} shows the ECD predictions for existing pharmaceutical molecules. 
We first visualize the ECD predictions for R/S type of hydroxybrevianamide~\cite{xu202117}, a natural product in Aspergillus sp. fungus. Fig.~\ref{fig:single_drug}(top) shows that ECDFormer can successfully predict the ECD spectra for R-type and S-type molecule pairs. We also visualize the ECD predictions for other pharmaceutical molecules, including Wulfenioidins.L~\cite{Tu2023}~(Anti-Zika Virus Effect), Purpurascenines.B~\cite{lam2023purpurascenines}~(Antagonist Effect), and Alkaloids~\cite{liu2023anti}~(Anti-inflammatory Effect). Our ECDFormer predictions also match the ECD theoretical spectra of these complex natural products with pharmaceutical effects.

%% file: Content/4discussion.tex
\section{Discussion}

This study proposes a research framework for integrating deep learning techniques into the field of chemistry to improve the efficiency of researchers in acquiring the ECD spectra of chiral molecules. The proposed ECDFormer focuses on several core issues including data collection, 3D characterization of chiral molecules, and understanding of chirality. Firstly, as the ECD spectra of each molecule are calculated consistently, this study mainly employs Python scripts for batch processing as well as generation of the data, thus providing a standardized CMCDS dataset. Secondly, a specialized neural network, ECDFormer, was established, and experimental results showed that it can directly obtain ECD spectra from the smiles of chiral molecules.

% \subsection{Limitations and Future Works}

% The experimental validation shows the ECDFormer model possesses a satisfactory ability to predict the ECD spectra of organic small molecules. At present, there are still some shortcomings in this study, which can be improved in the future. Firstly, when constructing the large-scale data of ECD spectra, the calculation omitted the conformational search of each molecule, because it could greatly increase the time and cost. This leads to some bias in the final ECD spectral data. Secondly, the basis group employed in DFT calculation restricts the range of available chiral molecules, which should not have elements heavier than iodine. Thirdly, all the molecules have only one chiral center, and those with multiple chiral centers were deliberated omitted. Despite these limitations, we still believe that the ECDFormer model holds great potential in determining the absolute configuration of chiral molecules and that the ECD spectra can be quickly obtained directly from the SMILES of the molecules.

The ECDFormer model's experimental validation demonstrates its proficient capability and generalization ability in predicting ECD spectra for small organic molecules, including single-chiral-centered molecules and multi-chiral-centered molecules. However, there are areas for improvement in this study that could be addressed in future research. Initially, in compiling the extensive ECD spectral data, we bypassed the conformational search for each molecule to minimize time and cost, which may have introduced some inaccuracies in the spectral data. Additionally, the choice of basis set in DFT calculations limits the spectrum of chiral molecules we can study, particularly excluding those containing elements heavier than iodine. Moreover, our focus was solely on molecules with a single chiral center, intentionally excluding those with multiple chiral centers. Despite these constraints, we remain optimistic about the ECDFormer model's potential in accurately determining the absolute configuration of chiral molecules. The model offers a rapid way to acquire ECD spectra directly from the SMILES notation of the molecules.

%% file: Content/5methods.tex
\section{Methods}

% \subsection{Construction of the CMCDS dataset}

\subsection{Problem Definition and Preliminary for Electronic Circular Dichroism Prediction}

We first briefly introduce the problem definition of the ECD prediction task for the convenience of description and discussion.

\subsubsection{Electronic Circular Dichroism Prediction Task. }
Generally, each chemical molecule has its electronic circular dichroism~(ECD). For molecule $M_{1}$, we represent the ECD of $M_{1}$ as $\{\mathcal{S}_{1:i}\}_{i=1}^{N_{w}}$, where $\mathcal{S}_{1:i}$ is the input light wavelength from 80 to 450nm, and $N_{w}$ is the ECD range from -200Mdeg to 200Mdeg. For $M_{1}$'s chiral-form molecule $\widetilde{M}_{1}$, we represent its ECD as $\{-\mathcal{S}_{1:i}\}_{i=1}^{N_{w}}$. When applying deep learning models for ECD prediction, a direct thought is to establish a site-level sequence prediction model to predict every $\mathcal{S}_{1:i}$ of the ECD. However, in practice analysis, molecular representation lacks the knowledge to reconstruct the site-level ECD sequence. Thus, we simplify the ECD prediction task from the chemical perspective, focusing on the peak features in the ECD sequence. Specifically, we represent molecule $M_{1}$'s ECD sequence as $\{\mathcal{P}_{1:j}\}_{j=1}^{N_{p}}$, where $\mathcal{P}_{1:j}$ is the j-th peak in ECD sequence, and $N_{p}$ is the peak number. The ECD prediction task aims to predict the peak number $N_{p}$, the peak position and height for $\mathcal{P}_{1:j}$. Under the new task setting, deep learning models achieve better performance on ECD prediction and chiral molecule distinguishment.

% 通常来说，每个化学分子都有它独特的ECD光谱。我们将化学分子M_i的ECD光谱表示为(公式XXXX)。 对于分子M_i对应的手性分子M_i, 他的ECD光谱完全相反。 对于深度学习处理ECD光谱预测任务，一个通常的想法是让模型预测整个ECD光谱，即 f_theta公式. 然而ECD光谱预测任务不会as simple as ..., in practical analysis, 由于存在峰宽度等随机变量，还原一整个序列过于困难. Thus, 我们从化学角度进行简化, 聚焦在光谱的核心元素： 峰, 并将ECD光谱预测任务从序列预测转换为峰特征预测，包括峰的数量, 位置, 正负。 在新的任务设定下模型取得更好的预测性能并更能区分手性分子

\subsubsection{Deep Leaning Models: Graph Neural Network~(GNN) and Transformer Network. }
% 中文注释加一些，引用一些AI论文
% Graph Convolutional Network~\cite{}属于图表征学习模型, 
% 由于分子很容易转换成图结构, GCN成为提取化学分子表征的常见模型~\cite{A, B, C}. 在输入graph后, GCN通过iteritavely 使用图卷积层来学习每个节点和它邻接节点矩阵的特征, 通过迭代化的学习, GCN能够关注graph的重点区域并学习良好的graph表征
Graph Neural Network~\cite{zhang2019graph} is an outstanding model for graph representation learning. For molecules, the atoms and chemical bonds are easy to interpret as a graph. Thus, GNN becomes the regular model to extract molecular representation~\cite{xu2023retention,mahmood2021masked,zhong2023retrosynthesis}. For an input molecular graph, GNN takes the weighted average of node features and their neighbor features, resulting in new representations for nodes. GNN iterates this process through multiple layers of fully connected layers to progressively propagate and aggregate information from the nodes, leading to richer molecule representations.

% Transformer是深度学习中处理序列的经典模型, 它使用堆叠的注意力模块来实现不同位置的序列特征的特征融合, 从而实现更好的序列预测. Specifically, Transformer分成 Encoder, Decoder两个部分, Encoder部分采用双向注意力模块, 旨在更好的融合输入特征, 而Decoder部分采用单向注意力模块, 旨在预测序列. 如上我们已经将ECD预测任务从简单的序列预测重新定义为峰值信息预测, 因此我们选用Transformer Encoder结构为其更强的特征融合能力

The Transformer~\cite{han2022survey} is a seminal deep-learning model for the sequence processing task. It utilizes stacked attention~\cite{vaswani2017attention} modules to fuse sequence features from different positions, thereby achieving improved sequence prediction performance. The Transformer contains two parts: the Encoder and the Decoder. The Encoder employs bidirectional attention modules to better integrate input features, while the Decoder employs unidirectional attention modules for sequence prediction. In this work, we have redefined the ECD prediction task as peak information prediction, and therefore, we apply the Transformer Encoder structure for its enhanced feature fusion capability.

\subsection{The Framework of the proposed ECDFormer}
The overview of our ECDFormer is illustrated in Fig.~\ref{fig:framework}. Our ECDFormer contains four major modules: (1).~Feature Extraction Module with GeoGNN~\cite{fang2022geometry,peng2020enhanced}, (2).~Peak Property Learning Module with Transformer Encoder, (3).~Peak Property Prediction Module, (4).~ECD Rendering Module. The workflow of our ECDFormer is described below.
% 加上一个光谱渲染模块, 将节点信息转换成可读的伪光谱序列

The Feature Extraction Module utilizes GeoGNN containing two graph convolutional networks to extract the molecule's geometric and descriptor information from the molecule's atom-bond graph and bond-angle graph. 
Then, the molecule representation features are input into the Peak Property Learning Module together with empty query tokens. With the transformer encoder structure, the Peak Property Learning Module extracts the peak-related features from the molecule features to the empty query tokens. 
In the Peak Property Prediction Module, the resulting peak-related features are simultaneously fed into three specific task heads: the peak-number head, the peak-position head, and the peak-height head to predict the peak properties. 
Finally, the ECD Rendering Module reconstructs the ECD spectra from the peak properties employing mathematical simulation methods. 
We further introduce more details about the Feature Extraction Module, Peak Property Learning Module, Peak Property Prediction Module, and ECD Rendering Module in the following subsections.

\subsection{Molecular Feature Extraction Module}
As shown in Fig.~\ref{fig:framework}, for the molecular feature extraction module, we apply the GeoGNN structure to encode molecular geometric features by modeling the atom-bond-angle corresponding relations. Compared with the traditional GNNs that only consider the atom-bond relationship, GeoGNN~\cite{fang2022geometry,peng2020enhanced} has a stronger ability in molecular representation modeling. 

Specifically, for an input molecule $M$, we denote its atom set as $\mathcal{V}$, its bond set as $\mathcal{E}$, and its bond-angle set as $\mathcal{A}$. Then we introduce $M$'s atom-bond graph $G$ and bond-angle graph $H$. The atom-bond graph is defined as $G=(\mathcal{V},\mathcal{E})$, where atom $u\in \mathcal{V}$ is regarded as the node of $G$ and bond $(u,v)\in \mathcal{E}$ as the edge of $G$. Similarly, the bond-angle graph is defined as $H=(\mathcal{E}, \mathcal{A})$, where bond $(u,v)\in \mathcal{E}$ is regarded as the node of $H$ and bond angle $(u,v,w)\in \mathcal{A}$ as the edge of $H$. Both the atom-bond graph and the bond-angle graph are input into the GeoGNN for further feature extraction.

Then, our feature extraction module learns the representation of atoms and bonds iteratively. For the k-th iteration, we use ${\rm\textbf{h}}_{u}$ and ${\rm\textbf{h}}_{uv}$ as the representation of atom $u$ and bond $(u,v)$.
To achieve information aggregation between the atom-bond graph $G$ and the bond-angle graph $H$, the representation vectors of the bonds are taken as the information link between $G$ and $H$. Specifically, the iteration of our feature extraction module contains two stages: 

In the first stage, the bonds’ representation vectors are learned by aggregating messages from the neighboring bonds and corresponding bond angles in the bond–angle graph $H$. Given bond $(u,v)$, in $k$-th iteration, its representation ${\rm \textbf{h}}_{uv}^{(k)}$ is formalized by: 
\begin{gather}
    {\rm \textbf{a}}_{uv}^{(k)}= \mathcal{F}_{bond-angle}^{(k)}(\{ ({\rm \textbf{h}}_{uv}^{(k-1)}, {\rm \textbf{h}}_{uw}^{(k-1)}, {\rm \textbf{x}}_{wuv}): w\in\mathcal{N}(u) \} \cup \{ ({\rm \textbf{h}}_{uv}^{(k-1)}, {\rm \textbf{h}}_{vw}^{(k-1)}, {\rm \textbf{x}}_{uvw}): w\in\mathcal{N}(v) \} ), \\
    {\rm \textbf{h}}_{uv}^{(k)} = \mathcal{W}_s*{\rm \textbf{h}}_{uv}^{(k-1)} + {\rm \textbf{a}}_{uv}^{(k)},
\end{gather}
where $\mathcal{N}(u)$ and $\mathcal{N}(v)$ are the neighbor atoms of $u$ and $v$. $\{ (u,w): w\in\mathcal{N}(u) \}\cup\{ (v,w): w\in\mathcal{N}(v) \}$ are the neighbor bonds of bond $(u,v)$. $\mathcal{F}_{bond-angle}$ is an MLP with two linear layers, acting as the message aggregation function. ${\rm \textbf{a}}_{uv}$ is bond $(u,v)$'s aggregated feature from neighbor bonds. Then, the bond $(u,v)$'s representation vector is updated according to ${\rm \textbf{a}}_{uv}$ in Eq.2.

In the second stage, with the updated bond representation in $H$, we further learn the atoms’ representation by aggregating messages from the neighboring atoms and the corresponding bond representations from $H$. Given an atom $u$, its representation ${\rm \textbf{h}}_{u}^{(k)}$ in the $k$-th iteration is formalized as: 
\begin{gather}
    {\rm \textbf{a}}_{u}^{(k)}= \mathcal{F}_{atom-bond}^{(k)}(\{ ({\rm \textbf{h}}_{u}^{(k-1)}, {\rm \textbf{h}}_{v}^{(k-1)}, {\rm \textbf{h}}_{uv}^{(k-1)}): v\in\mathcal{N}(u) \} ), \\
    {\rm \textbf{h}}_{u}^{(k)} = \mathcal{W}_s*{\rm \textbf{h}}_{u}^{(k-1)} + {\rm \textbf{a}}_{u}^{(k)},
\end{gather}
where $\mathcal{N}(u)$ represents the neighbor atoms of atom $u$. $\mathcal{F}_{atom-bond}$ is an MLP with two linear layers, acting as the message aggregation function. ${\rm \textbf{a}}_{u}$ is the atom $u$'s aggregated feature from neighbor atoms. The representation of $u$ is updated according to ${\rm \textbf{a}}_{u}$ in Eq.4.

After all iterations, we calculate the molecular global representation ${\rm \textbf{h}}_G$ by summarizing and pooling over the atoms' representation $\{{\rm \textbf{h}}_{u}\}, \forall u\in\mathcal{V}$. We further take the input molecule $M$'s global representation ${\rm \textbf{h}}_G$ and atom representations $\{{\rm \textbf{h}}_{u}\}$ as the input of further modules.

\subsection{Peak Property Learning Module}
Our peak property learning module aims to fuse the atom representation features and extract the key features with peak property information. We apply the transformer encoder as the fusion model due to its powerful feature fusion capability enabled by its cross-attention structure. Specifically, for molecule $M$, we first random initialize a set of tokens $\{ {\rm \textbf{Q}}_i \}_{i=1}^n$ as the peak tokens for $M$. Then we combine the peak tokens $\{ {\rm \textbf{Q}}_i \}$ with $M$'s global feature ${\rm \textbf{h}}_G$ and $M$'s atom features $\{{\rm \textbf{h}}_{u}\}$ as the input tokens for transformer encoder, which is formalized as: 
\begin{gather}
    [{\rm \textbf{Q}}_j, {\rm \textbf{h}}_{G, j}, {\rm \textbf{h}}_{u,j}] = {{Layer}_j}({\rm \textbf{Q}}_{j-1}, {\rm \textbf{h}}_{G, j-1}, {\rm \textbf{h}}_{u,j-1})
\end{gather}
where $Layer_j$ represents the $j$-th transformer encoder layer. After the cross attention in $N$ transformer encoder layers, the ${\rm \textbf{Q}}_N$ denote the final representations for peak tokens. We get the peak number $N_p$ from the molecular ground-truth ECD spectra $\{\mathcal{P}_{1:j}\}_{j=1}^{N_{p}}$. Thus, we extract the first $N_p$ peak token features from $\{ {\rm \textbf{Q}}_i \}_{i=1}^n$ as the peak property information of molecule $M$.

\subsection{Peak Property Prediction Module}
Our peak property prediction module aims to reconstruct the peak property, including the peak number, peak symbol, and peak position, from the output features of the peak property learning module. For the peak number prediction, we apply the two-layer MLP to predict the peak number from the molecule global feature ${\rm \textbf{h}}_{G}$, which is formalized as:
\begin{gather}
    \mathcal{P}_{num} = {\rm Linear} ({\rm ReLU}( {\rm Linear}( {\rm \textbf{h}}_{G} )))
\end{gather}
where $\mathcal{P}_{num}$ represents the peak number for the ecd spectra of molecule $M$. For peak height $\mathcal{P}_{height}$ and peak position $\mathcal{P}_{pos}$. we also apply two separate two-layer MLPs to predict $\mathcal{P}_{symbol}$ and $\mathcal{P}_{pos}$ from the corresponding peak token ${\rm \textbf{Q}}_i$ from $\{ {\rm \textbf{Q}}_i \}_{i=1}^n$. 
\begin{gather}
    \mathcal{P}_{pos} = {\rm Linear} ({\rm ReLU}( {\rm Linear}( {\rm \textbf{Q}}_i ))), \quad
    \mathcal{P}_{symbol} = {\rm Linear} ({\rm ReLU}( {\rm Linear}( {\rm \textbf{Q}}_i ))), 
\end{gather}
Here we predict all three peak properties that are vital for ECD spectra prediction.

\subsection{ECD Spectra Rendering Module}
The final module, the ECD spectra rendering module, aims to render the predicted ECD spectra from the abstract peak properties. 
We employ the Gaussian noise distribution model to fit the spectral curve. Specifically, given the position $l_p$ and the corresponding height $l_h$ of a peak, we set a Gaussian noise distribution with mean value: $\mu=l_p$ and a standard deviation of $\sigma=l_h$. We then extract the distribution range of $[\mu-6\sigma, \mu+6\sigma]$ as the fitting curve for the peak.
We render the predicted ECD spectra for molecule $M$ by combining the fitting curves of all predicted peaks.

\subsection{Experimental Settings and Training Hyperparameters}
In the molecular feature extraction module, we set the number of GINConv in GeoGNN to be $5$ and the graph pooling strategy to be $summation$. The embedding dimension of molecular features is $128$ and the batch size for ECDFormer is $256$. We apply the AdamW~\cite{kingma2014adam} optimizer implemented in Pytorch. The learning rate $= 1e^{-3}$. For better convergence, we apply the $StepLR$ schedular with a decreasing rate $= 0.25$ to adaptively adjust the learning rate. During training, the CMCDS dataset is randomly divided into $90/5/5$ for train/valid/test splits. ECDFormer is trained with 1000 epochs, selecting the best valid checkpoint for the testing procedure. For other deep-learning baselines, we apply the learning rate $=5e^{-4}$ and epoch $=1500$, while other parameters are the same as ECDFormer.

%% file: Content/6DataAvailable.tex
\section*{Data Availability}

The first large-scale ECD spectra dataset for chiral molecules, the \textbf{CMCDS} dataset, has been deposited in the Github repository, \href{https://github.com/HowardLi1984/ECDFormer}{my-dataset-link}.

%% file: Content/7CodeAvailable.tex
\section*{Code Availability}

All code used in data analysis and preparation of the manuscript, alongside a description of necessary steps for reproducing results, can be found in a GitHub repository accompanying this manuscript: \href{https://github.com/HowardLi1984/ECDFormer}{my-github-link}.

%% file: main.bbl
\begin{thebibliography}{10}
\urlstyle{rm}
\expandafter\ifx\csname url\endcsname\relax
  \def\url#1{\texttt{#1}}\fi
\expandafter\ifx\csname urlprefix\endcsname\relax\def\urlprefix{URL }\fi
\expandafter\ifx\csname doiprefix\endcsname\relax\def\doiprefix{DOI: }\fi
\providecommand{\bibinfo}[2]{#2}
\providecommand{\eprint}[2][]{\url{#2}}

\bibitem{noyori2002asymmetric}
\bibinfo{author}{Noyori, R.}
\newblock \bibinfo{journal}{\bibinfo{title}{Asymmetric catalysis: science and opportunities (nobel lecture)}}.
\newblock {\emph{\JournalTitle{Angewandte Chemie International Edition}}} \textbf{\bibinfo{volume}{41}}, \bibinfo{pages}{2008--2022} (\bibinfo{year}{2002}).

\bibitem{list20212021}
\bibinfo{author}{List, B.} \& \bibinfo{author}{MacMillan, D.}
\newblock \bibinfo{journal}{\bibinfo{title}{The 2021 nobel prize in chemistry: asymmetric catalysis with small organic molecules}}.
\newblock {\emph{\JournalTitle{Current Science}}} \textbf{\bibinfo{volume}{121}}, \bibinfo{pages}{1148} (\bibinfo{year}{2021}).

\bibitem{amabilino2006supramolecular}
\bibinfo{author}{Amabilino, D.~B.} \& \bibinfo{author}{Veciana, J.}
\newblock \bibinfo{journal}{\bibinfo{title}{Supramolecular chiral functional materials}}.
\newblock {\emph{\JournalTitle{Supramolecular Chirality}}} \bibinfo{pages}{253--302} (\bibinfo{year}{2006}).

\bibitem{shen2020supramolecular}
\bibinfo{author}{Shen, B.}, \bibinfo{author}{Kim, Y.} \& \bibinfo{author}{Lee, M.}
\newblock \bibinfo{journal}{\bibinfo{title}{Supramolecular chiral 2d materials and emerging functions}}.
\newblock {\emph{\JournalTitle{Advanced Materials}}} \textbf{\bibinfo{volume}{32}}, \bibinfo{pages}{1905669} (\bibinfo{year}{2020}).

\bibitem{teng2022advances}
\bibinfo{author}{Teng, Y.} \emph{et~al.}
\newblock \bibinfo{journal}{\bibinfo{title}{Advances and applications of chiral resolution in pharmaceutical field}}.
\newblock {\emph{\JournalTitle{Chirality}}} \textbf{\bibinfo{volume}{34}}, \bibinfo{pages}{1094--1119} (\bibinfo{year}{2022}).

\bibitem{lininger2023chirality}
\bibinfo{author}{Lininger, A.} \emph{et~al.}
\newblock \bibinfo{journal}{\bibinfo{title}{Chirality in light--matter interaction}}.
\newblock {\emph{\JournalTitle{Advanced Materials}}} \textbf{\bibinfo{volume}{35}}, \bibinfo{pages}{2107325} (\bibinfo{year}{2023}).

\bibitem{evers2022theory}
\bibinfo{author}{Evers, F.} \emph{et~al.}
\newblock \bibinfo{journal}{\bibinfo{title}{Theory of chirality induced spin selectivity: Progress and challenges}}.
\newblock {\emph{\JournalTitle{Advanced Materials}}} \textbf{\bibinfo{volume}{34}}, \bibinfo{pages}{2106629} (\bibinfo{year}{2022}).

\bibitem{zhang2019great}
\bibinfo{author}{Zhang, W.} \emph{et~al.}
\newblock \bibinfo{journal}{\bibinfo{title}{Great concern for chiral pharmaceuticals from the thalidomide tragedy}}.
\newblock {\emph{\JournalTitle{Univ. Chem}}} \textbf{\bibinfo{volume}{34}}, \bibinfo{pages}{1--12} (\bibinfo{year}{2019}).

\bibitem{ebeling2018assigning}
\bibinfo{author}{Ebeling, D.} \emph{et~al.}
\newblock \bibinfo{journal}{\bibinfo{title}{Assigning the absolute configuration of single aliphatic molecules by visual inspection}}.
\newblock {\emph{\JournalTitle{Nature communications}}} \textbf{\bibinfo{volume}{9}}, \bibinfo{pages}{2420} (\bibinfo{year}{2018}).

\bibitem{menna2019challenges}
\bibinfo{author}{Menna, M.}, \bibinfo{author}{Imperatore, C.}, \bibinfo{author}{Mangoni, A.}, \bibinfo{author}{Della~Sala, G.} \& \bibinfo{author}{Taglialatela-Scafati, O.}
\newblock \bibinfo{journal}{\bibinfo{title}{Challenges in the configuration assignment of natural products. a case-selective perspective}}.
\newblock {\emph{\JournalTitle{Natural product reports}}} \textbf{\bibinfo{volume}{36}}, \bibinfo{pages}{476--489} (\bibinfo{year}{2019}).

\bibitem{junior2023absolute}
\bibinfo{author}{Junior, F. M. d.~S.} \& \bibinfo{author}{Junior, J. M.~B.}
\newblock \bibinfo{journal}{\bibinfo{title}{Absolute configuration from chiroptical spectroscopy}}.
\newblock {\emph{\JournalTitle{Chiral Separations and Stereochemical Elucidation: Fundamentals, Methods, and Applications}}} \bibinfo{pages}{551--591} (\bibinfo{year}{2023}).

\bibitem{janet2020machine}
\bibinfo{author}{Janet, J.~P.} \& \bibinfo{author}{Kulik, H.~J.}
\newblock \emph{\bibinfo{title}{Machine Learning in chemistry}}, vol.~\bibinfo{volume}{1} (\bibinfo{publisher}{American Chemical Society}, \bibinfo{year}{2020}).

\bibitem{de2019synthetic}
\bibinfo{author}{de~Almeida, A.~F.}, \bibinfo{author}{Moreira, R.} \& \bibinfo{author}{Rodrigues, T.}
\newblock \bibinfo{journal}{\bibinfo{title}{Synthetic organic chemistry driven by artificial intelligence}}.
\newblock {\emph{\JournalTitle{Nature Reviews Chemistry}}} \textbf{\bibinfo{volume}{3}}, \bibinfo{pages}{589--604} (\bibinfo{year}{2019}).

\bibitem{hermann2022ab}
\bibinfo{author}{Hermann, J.} \emph{et~al.}
\newblock \bibinfo{journal}{\bibinfo{title}{Ab-initio quantum chemistry with neural-network wavefunctions}}.
\newblock {\emph{\JournalTitle{arXiv preprint arXiv:2208.12590}}}  (\bibinfo{year}{2022}).

\bibitem{xu2023retention}
\bibinfo{author}{Xu, H.}, \bibinfo{author}{Lin, J.}, \bibinfo{author}{Zhang, D.} \& \bibinfo{author}{Mo, F.}
\newblock \bibinfo{journal}{\bibinfo{title}{Retention time prediction for chromatographic enantioseparation by quantile geometry-enhanced graph neural network}}.
\newblock {\emph{\JournalTitle{Nature Communications}}} \textbf{\bibinfo{volume}{14}}, \bibinfo{pages}{3095} (\bibinfo{year}{2023}).

\bibitem{fang2022geometry}
\bibinfo{author}{Fang, X.} \emph{et~al.}
\newblock \bibinfo{journal}{\bibinfo{title}{Geometry-enhanced molecular representation learning for property prediction}}.
\newblock {\emph{\JournalTitle{Nature Machine Intelligence}}} \textbf{\bibinfo{volume}{4}}, \bibinfo{pages}{127--134} (\bibinfo{year}{2022}).

\bibitem{vaswani2017attention}
\bibinfo{author}{Vaswani, A.} \emph{et~al.}
\newblock \bibinfo{journal}{\bibinfo{title}{Attention is all you need}}.
\newblock {\emph{\JournalTitle{Advances in neural information processing systems}}} \textbf{\bibinfo{volume}{30}} (\bibinfo{year}{2017}).

\bibitem{g16}
\bibinfo{author}{Frisch, M.~J.} \emph{et~al.}
\newblock \bibinfo{title}{Gaussian˜16 {R}evision {B}.01} (\bibinfo{year}{2016}).
\newblock \bibinfo{note}{Gaussian Inc. Wallingford CT}.

\bibitem{stephens1994ab}
\bibinfo{author}{Stephens, P.~J.}, \bibinfo{author}{Devlin, F.~J.}, \bibinfo{author}{Chabalowski, C.~F.} \& \bibinfo{author}{Frisch, M.~J.}
\newblock \bibinfo{journal}{\bibinfo{title}{Ab initio calculation of vibrational absorption and circular dichroism spectra using density functional force fields}}.
\newblock {\emph{\JournalTitle{The Journal of physical chemistry}}} \textbf{\bibinfo{volume}{98}}, \bibinfo{pages}{11623--11627} (\bibinfo{year}{1994}).

\bibitem{yanai2004new}
\bibinfo{author}{Yanai, T.}, \bibinfo{author}{Tew, D.~P.} \& \bibinfo{author}{Handy, N.~C.}
\newblock \bibinfo{journal}{\bibinfo{title}{A new hybrid exchange-correlation functional using the coulomb-attenuating method (cam-b3lyp)}}.
\newblock {\emph{\JournalTitle{Chemical physics letters}}} \textbf{\bibinfo{volume}{393}}, \bibinfo{pages}{51--57} (\bibinfo{year}{2004}).

\bibitem{zou2023deep}
\bibinfo{author}{Zou, Z.} \emph{et~al.}
\newblock \bibinfo{journal}{\bibinfo{title}{A deep learning model for predicting selected organic molecular spectra}}.
\newblock {\emph{\JournalTitle{Nature Computational Science}}} \bibinfo{pages}{1--8} (\bibinfo{year}{2023}).

\bibitem{rogers2019electronic}
\bibinfo{author}{Rogers, D.~M.} \emph{et~al.}
\newblock \bibinfo{journal}{\bibinfo{title}{Electronic circular dichroism spectroscopy of proteins}}.
\newblock {\emph{\JournalTitle{Chem}}} \textbf{\bibinfo{volume}{5}}, \bibinfo{pages}{2751--2774} (\bibinfo{year}{2019}).

\bibitem{mao2023crossentropy}
\bibinfo{author}{Mao, A.}, \bibinfo{author}{Mohri, M.} \& \bibinfo{author}{Zhong, Y.}
\newblock \bibinfo{journal}{\bibinfo{title}{Cross-entropy loss functions: Theoretical analysis and applications}}.
\newblock {\emph{\JournalTitle{arXiv preprint arXiv:2304.07288}}}  (\bibinfo{year}{2023}).

\bibitem{nagy2019sesca}
\bibinfo{author}{Nagy, G.}, \bibinfo{author}{Igaev, M.}, \bibinfo{author}{Jones, N.~C.}, \bibinfo{author}{Hoffmann, S.~V.} \& \bibinfo{author}{Grubmuller, H.}
\newblock \bibinfo{journal}{\bibinfo{title}{Sesca: predicting circular dichroism spectra from protein molecular structures}}.
\newblock {\emph{\JournalTitle{Journal of chemical theory and computation}}} \textbf{\bibinfo{volume}{15}}, \bibinfo{pages}{5087--5102} (\bibinfo{year}{2019}).

\bibitem{micsonai2021bestsel}
\bibinfo{author}{Micsonai, A.}, \bibinfo{author}{Buly{\'a}ki, {\'E}.} \& \bibinfo{author}{Kardos, J.}
\newblock \bibinfo{journal}{\bibinfo{title}{Bestsel: from secondary structure analysis to protein fold prediction by circular dichroism spectroscopy}}.
\newblock {\emph{\JournalTitle{Structural Genomics: General Applications}}} \bibinfo{pages}{175--189} (\bibinfo{year}{2021}).

\bibitem{zhao2021accurate}
\bibinfo{author}{Zhao, L.} \emph{et~al.}
\newblock \bibinfo{journal}{\bibinfo{title}{Accurate machine learning prediction of protein circular dichroism spectra with embedded density descriptors}}.
\newblock {\emph{\JournalTitle{JACS Au}}} \textbf{\bibinfo{volume}{1}}, \bibinfo{pages}{2377--2384} (\bibinfo{year}{2021}).

\bibitem{artrith2021best}
\bibinfo{author}{Artrith, N.} \emph{et~al.}
\newblock \bibinfo{journal}{\bibinfo{title}{Best practices in machine learning for chemistry}}.
\newblock {\emph{\JournalTitle{Nature chemistry}}} \textbf{\bibinfo{volume}{13}}, \bibinfo{pages}{505--508} (\bibinfo{year}{2021}).

\bibitem{wei2019machine}
\bibinfo{author}{Wei, J.} \emph{et~al.}
\newblock \bibinfo{journal}{\bibinfo{title}{Machine learning in materials science}}.
\newblock {\emph{\JournalTitle{InfoMat}}} \textbf{\bibinfo{volume}{1}}, \bibinfo{pages}{338--358} (\bibinfo{year}{2019}).

\bibitem{shi2015convolutional}
\bibinfo{author}{Shi, X.} \emph{et~al.}
\newblock \bibinfo{journal}{\bibinfo{title}{Convolutional lstm network: A machine learning approach for precipitation nowcasting}}.
\newblock {\emph{\JournalTitle{Advances in neural information processing systems}}} \textbf{\bibinfo{volume}{28}} (\bibinfo{year}{2015}).

\bibitem{chung2014empirical}
\bibinfo{author}{Chung, J.}, \bibinfo{author}{Gulcehre, C.}, \bibinfo{author}{Cho, K.} \& \bibinfo{author}{Bengio, Y.}
\newblock \bibinfo{journal}{\bibinfo{title}{Empirical evaluation of gated recurrent neural networks on sequence modeling}}.
\newblock {\emph{\JournalTitle{arXiv preprint arXiv:1412.3555}}}  (\bibinfo{year}{2014}).

\bibitem{xu202117}
\bibinfo{author}{Xu, W.-F.} \emph{et~al.}
\newblock \bibinfo{journal}{\bibinfo{title}{17-hydroxybrevianamide n and its n1-methyl derivative, quinazolinones from a soft-coral-derived aspergillus sp. fungus: 13 s enantiomers as the true natural products}}.
\newblock {\emph{\JournalTitle{Journal of Natural Products}}} \textbf{\bibinfo{volume}{84}}, \bibinfo{pages}{1353--1358} (\bibinfo{year}{2021}).

\bibitem{Tu2023}
\bibinfo{author}{Tu, W.-C.} \emph{et~al.}
\newblock \bibinfo{journal}{\bibinfo{title}{Wulfenioidins d–n, structurally diverse diterpenoids with anti-zika virus activity isolated from orthosiphon wulfenioides}}.
\newblock {\emph{\JournalTitle{Journal of Natural Products}}} \textbf{\bibinfo{volume}{86}}, \bibinfo{pages}{2348--2359} (\bibinfo{year}{2023}).

\bibitem{lam2023purpurascenines}
\bibinfo{author}{Lam, Y.~T.} \emph{et~al.}
\newblock \bibinfo{journal}{\bibinfo{title}{Purpurascenines a--c, azepino-indole alkaloids from cortinarius purpurascens: Isolation, biosynthesis, and activity studies on the 5-ht2a receptor}}.
\newblock {\emph{\JournalTitle{Journal of Natural Products}}}  (\bibinfo{year}{2023}).

\bibitem{liu2023anti}
\bibinfo{author}{Liu, F.} \emph{et~al.}
\newblock \bibinfo{journal}{\bibinfo{title}{Anti-inflammatory quinoline alkaloids from the roots of waltheria indica}}.
\newblock {\emph{\JournalTitle{Journal of Natural Products}}} \textbf{\bibinfo{volume}{86}}, \bibinfo{pages}{276--289} (\bibinfo{year}{2023}).

\bibitem{zhang2019graph}
\bibinfo{author}{Zhang, S.}, \bibinfo{author}{Tong, H.}, \bibinfo{author}{Xu, J.} \& \bibinfo{author}{Maciejewski, R.}
\newblock \bibinfo{journal}{\bibinfo{title}{Graph convolutional networks: a comprehensive review}}.
\newblock {\emph{\JournalTitle{Computational Social Networks}}} \textbf{\bibinfo{volume}{6}}, \bibinfo{pages}{1--23} (\bibinfo{year}{2019}).

\bibitem{mahmood2021masked}
\bibinfo{author}{Mahmood, O.}, \bibinfo{author}{Mansimov, E.}, \bibinfo{author}{Bonneau, R.} \& \bibinfo{author}{Cho, K.}
\newblock \bibinfo{journal}{\bibinfo{title}{Masked graph modeling for molecule generation}}.
\newblock {\emph{\JournalTitle{Nature communications}}} \textbf{\bibinfo{volume}{12}}, \bibinfo{pages}{3156} (\bibinfo{year}{2021}).

\bibitem{zhong2023retrosynthesis}
\bibinfo{author}{Zhong, W.}, \bibinfo{author}{Yang, Z.} \& \bibinfo{author}{Chen, C. Y.-C.}
\newblock \bibinfo{journal}{\bibinfo{title}{Retrosynthesis prediction using an end-to-end graph generative architecture for molecular graph editing}}.
\newblock {\emph{\JournalTitle{Nature Communications}}} \textbf{\bibinfo{volume}{14}}, \bibinfo{pages}{3009} (\bibinfo{year}{2023}).

\bibitem{han2022survey}
\bibinfo{author}{Han, K.} \emph{et~al.}
\newblock \bibinfo{journal}{\bibinfo{title}{A survey on vision transformer}}.
\newblock {\emph{\JournalTitle{IEEE transactions on pattern analysis and machine intelligence}}} \textbf{\bibinfo{volume}{45}}, \bibinfo{pages}{87--110} (\bibinfo{year}{2022}).

\bibitem{peng2020enhanced}
\bibinfo{author}{Peng, Y.} \emph{et~al.}
\newblock \bibinfo{journal}{\bibinfo{title}{Enhanced graph isomorphism network for molecular admet properties prediction}}.
\newblock {\emph{\JournalTitle{Ieee Access}}} \textbf{\bibinfo{volume}{8}}, \bibinfo{pages}{168344--168360} (\bibinfo{year}{2020}).

\bibitem{kingma2014adam}
\bibinfo{author}{Kingma, D.~P.} \& \bibinfo{author}{Ba, J.}
\newblock \bibinfo{journal}{\bibinfo{title}{Adam: A method for stochastic optimization}}.
\newblock {\emph{\JournalTitle{arXiv preprint arXiv:1412.6980}}}  (\bibinfo{year}{2014}).

\bibitem{olboyle2011confab}
\bibinfo{author}{OLBoyle, N.}, \bibinfo{author}{Vandermeersch, T.} \& \bibinfo{author}{Hutchison, G.}
\newblock \bibinfo{journal}{\bibinfo{title}{Confab-generation of diverse low energy conformers}}.
\newblock {\emph{\JournalTitle{Journal of Cheminformatics}}}  (\bibinfo{year}{2011}).

\end{thebibliography}
